\documentclass[10pt,letterpaper,twocolumn]{article} 

\usepackage{ol2}
\usepackage[draft]{hyperref}
\usepackage{amsmath}

\begin{document}

\twocolumn[ 

\title{Absolute calibration of photodetectors: photocurrent multiplication versus photocurrent subtraction}


\author{I.~N.~Agafonov$^1$, M.~V.~Chekhova$^{1,2}$, T.~Sh.~Iskhakov$^{2,*}$, A.~N.~Penin$^1$, G.~O.~Rytikov$^1$, and O.~A.~Shumilkina$^1$}

\address{
$^1$Physics Department, Moscow State University, \\ Leninskiye Gory 1-2, Moscow 119991, Russia
\\
$^2$Max-Planck Institute for the Science of Light, \\  Guenther-Scharowsky-Str. 1 / Bau 24, Erlangen  D-91058, Germany \\
$^*$Corresponding author: Timur.Iskhakov@mpl.mpg.de
}

\begin{abstract} We report testing of the new absolute method of photodetectors calibration based on the difference-signal measurement for two-mode squeezed vacuum by comparison with the traditional absolute method based on the coincidence counting. Using low-gain parametric down conversion we have measured the quantum efficiency of a counting detector by both methods. The difference-signal method was adapted for the counting detectors by taking into account the dead-time effect.

\end{abstract}


 ] 

\noindent Spontaneous parametric down conversion (SPDC) has important metrological applications. Perfect pairwise correlation of photons enables the absolute quantum efficiency measurement (\textit{absolute calibration})  of photon-counting detectors (also called single-photon detectors). This technique, known as the Klyshko method, is based on measuring the second-order intensity correlation function (CF). This is usually realized by measuring the rate of photocount coincidences for two detectors registering signal and idler photons of SPDC~\cite{Klyshko}. The method requires no reference sources or detectors, which makes it absolute. This technique was used many times~\cite{Malygin81,P.G.Kwiat,Novero} and it has been verified by the comparison with a reference detector~\cite{S.Polyakov}. However, it has a disadvantage: at high-gain PDC nonclassical intensity correlations between the signal and idler beams cannot be observed via CF measurement because of the high background of intensity CFs~\cite{NL,Brida2006}. For this reason, the method cannot be applied for the calibration of analog detectors.

However, there is another strategy for observing the signal-idler correlations, valid at any parametric gain (PG)~\cite{IskhakovJL}. This strategy is based on measuring the variance of the intensity difference. For PDC, photon-number fluctuations in the signal and idler beams are always identical, which is why they are called twin beams~\cite{Twinbeams}. The variance of their photon-number difference is reduced below the shot-noise level (SNL) and turns into zero in the absence of losses. Because this effect is extremely sensitive to the losses in the optical channel, including the non-ideal quantum efficiency of the detectors, it can form the base for another method of absolute calibration~\cite{Brida2006}. This method was applied for the calibration of an analog detector (CCD camera) ~\cite{Brida2010}; however it was not verified so far through the comparison with the conventional method.

In this Letter we present the first, to our knowledge, verification of the difference-signal method by comparing it with the Klyshko one. As both methods can be equally realized only at low PDC gain, a photon-counting detector was chosen for the test. Particular attention was paid to adapting the difference-signal measurement for counting detectors by taking into account the dead-time (DT) effect. It should be noted here that in both methods, it is not only the detector that is calibrated but the whole optical channel after the crystal.

Both methods considered in this Letter are based on the pairwise correlations in the PDC radiation. In the Klyshko method, the quantum efficiency $\eta_{i}$ of the calibrated (idler) detector can be found as the ratio of the coincidence number $N_c$ to the photocount number $N_s$ of the reference (signal) detector~\cite{P.G.Kwiat}
\begin{equation}
\eta_i=\frac{N_c-N_{ac}}{N_s-N_{bn}},
\label{coin}
\end{equation}
here $N_{ac}$ is the number of accidental coincidences and $N_{bn}$ is the background noise.

There are two conditions to be satisfied for the realization of this method. The first one is that the PG
should be low, so that the number of accidental coincidences is small. The second condition is that all detected modes in the reference channel should be covered by the conjugate modes in the calibrated one. Only then, a photodetection in the reference channel guarantees the existence of the photon at the input of the calibrated detector.

The difference-signal method is based on measuring the variance of the photocount number (photocurrent) difference
in the signal and idler beams, $N_-\equiv N_s-N_i$. The variance of $N_-$, normalized to the mean sum photocount number $\langle{N_+}\rangle$ in these channels, defines the \textit{noise reduction factor} (NRF)~\cite{Heidmann}: $NRF\equiv\mathrm{Var}(N_-)/\langle{N_+}\rangle$. For coherent radiation, $NRF=1$, which corresponds to the SNL. For squeezed-vacuum twin beams registered by two detectors with QEs equal to $\eta$, $NRF=1-\eta$. If signal and idler channels have different QEs $\eta_{s,i}$,
\begin{equation}
NRF=1-2\frac{\eta_i\eta_s}{\eta_i+\eta_s}+\langle N_+\rangle\frac{(\eta_i-\eta_s)^2}{(\eta_i+\eta_s)^2},
\label{NRFetadif}
\end{equation}
Therefore, unbalanced optical losses at high $\langle N_+\rangle$ lead to the growth of the difference-signal noise and, at some point, to the impossibility of the noise reduction below SNL~\cite{Agafonov2010}. To avoid this problem at high-gain  PDC one should manually balance the QEs by introducing additional losses into the optical channel or balance the signals in the channels numerically~\cite{Brida2010}.

Note that the high QE of the optical channel is not sufficient to guarantee a considerable reduction in the difference-signal noise for PDC in experiment. A very critical condition is conjugate multi-mode registration of signal and idler beams~\cite{Agafonov2010}.

Prior to applying the difference-signal method, one should adapt it to the photon-counting detectors, which mainly implies taking into account the dead-time effect. Each photodetection or a dark photocount blinds the avalanche photodiode for a period of time, which is called the dead time. The effect is especially noticeable when registering pulsed radiation with the pulse duration shorter than DT. Note that Eq.~(\ref{NRFetadif}) was obtained with the DT effect neglected, i.e, the pulse duration (or the measurement time) assumed to be much longer than the DT, so that the detector can register many photons within a single pulse. Now, let us consider the opposite case: the pulse duration is much shorter than the DT, and the  detector cannot register more then one photon during the pulse. Therefore, the fluctuations of photocount numbers become suppressed. For example, even for coherent radiation at the input of the detector one can observe sub-Poissonian statistics of photocounts. Using the simple Bernoulli model applied, for instance, in \cite{Steinberg}, and assuming multimode thermal statistics in each of the twin beams, we have found that the third term in the right-hand side of (\ref{NRFetadif}) should be replaced by
\begin{equation}
\Delta\equiv\langle N_+\rangle\left[\frac{\eta_i\eta_s}{\eta_i+\eta_s}- \frac{\eta_i^2+\eta_s^2+\eta_i^2\eta_s^2}{(\eta_i+\eta_s)^2}+\frac{2\eta_i^2\eta_s^2}{(\eta_i+\eta_s)^3}\right],
\label{NRFlarge}
\end{equation}
with $\langle N_+\rangle\ll1$. The DT correction converts (\ref{NRFetadif}) to a quadratic equation for $\eta_i$ whose positive root should be considered as the QE of the idler channel. The DT influence can be reduced by decreasing the signals but only at the cost of increasing the data acquisition time.

The experimental setup is shown in Fig. 1. PDC was exited by focused pulsed radiation of the third
harmonic of Nd:YAG laser with the wavelength 355 nm, pulse duration 5 ns, and repetition rate 10 kHz in a 3 mm $\mathrm{LiIO_3}$ crystal placed into the beam waist of diameter $0.4$ mm. The crystal was cut for type-I collinear phase matching, the signal and idler central wavelengths being $\lambda_s=650\,\hbox{nm},\,\lambda_i=780\,\hbox{nm}$, but could be tilted to change these wavelengths. After the crystal, the pump radiation was cut off by a dichroic mirror M and a red-glass filter RG. The signal and idler beams were separated by a dichroic beamsplitter DBS. The angular spectra registered at the DBS output were determined by two
iris apertures $A_1$,$A_2$. The lenses $L_1$ and $L_2$ focused the correlated photon beams on the
avalanche photodiodes operating in the Geiger mode ($D_1$,$D_2$). Signals from the detectors were analyzed
by the counters $N1$,$N2$ and a coincidence circuit (CC) with the time resolution 4.2 ns. Both counters
were gated by 30 ns pulses synchronized with the laser pulses, to suppress the influence of the dark counts. All the optical elements after the crystal had antireflection coating. The averaged single-photon count rates did not exceed $2\times10^{-2}$ per pulse but even then,
the DT effect was noticeable. Since the QE depends on the temperature of the detector and
the bias voltage above the breakdown, these parameters were kept at the constant level $T=-25 C$, $U=15 V$.
\begin{figure} [h]
\centerline{\includegraphics[width=8cm]{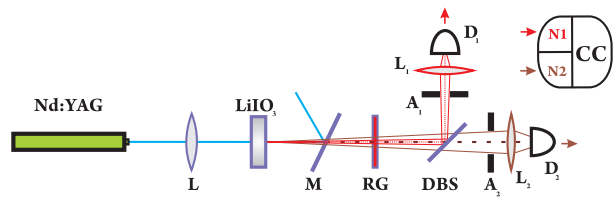}}
\caption{Experimental setup}
\end{figure}

In the Klyshko method, the frequency and angular spectra were restricted in the reference channel, by a spectral filter with 10 nm FWHM (IF) and a 2 mm aperture A2, while the calibrated channel had no filters and a larger aperture (8 mm). The quantum efficiency was calculated from (\ref{coin}). The number of accidental coincidences was calculated as $N_{ac}=N_1N_2K$, where $K=0.65$~\cite{O.A.Ivanova}.

In the difference-signal method, conjugate transverse mode registration was provided by choosing the diameters of $A_1$,$A_2$ as $D_s=5$  mm and $D_i=6$ mm, according to the condition~\cite{Agafonov2010}:
\begin{equation}
D_i/D_s={\lambda_i}/{\lambda_s}
\label{pm}
\end{equation}
The quantum efficiency $\eta_i$ of the calibrated detector was calculated from Eq.~(\ref{NRFetadif}), with the last term in the form of (\ref{NRFlarge}) and the ratio $\eta_i/\eta_s$ taken from the measured data.

In our comparative test, we measured the QE by both methods as a function of three parameters (Fig.2): losses in the calibrated channel, wavelength of the registered light, and the intensity of external light, which saturated the detector and hence reduced its QE. Fig. 2a shows the QE measured versus the optical channel transmission $T$. Additional losses were introduced by a polarization filter placed in the calibrated channel. The theoretical dependence (dashed line) is  $\eta=T\eta_0$, where $\eta_0$ is the arithmetic mean of the QE measured without additional losses by the difference-signal method, $\eta_D=0.256\pm0.004$, and the Klyshko one, $\eta_K=0.258\pm0.004$. Note that additional losses were only introduced into the calibrated channel, hence (\ref{NRFetadif}) was used in strongly asymmetric form, with $\eta_i\ll\eta_s$. Still, perfect agreement was obtained between the results of the two methods, which confirms the validity of the difference-signal one.

Figure 2(b) shows the QE dependence on the noise caused by an external continuous light source (a luminescent lamp with a variable intensity), which
reduced the QE due to the DT effect. The abscissas in Fig. 2(b) show the counting rate of the detector registering the external light in the non-gated mode, with the PDC blocked.
 It is important that although the QE was measured in the gated regime, the external noise still influenced all measured values in (\ref{coin})-(\ref{NRFlarge}). However, due to the statistical independence of the PDC radiation and the external light, the contribution of the latter was additive and could be measured separately and then subtracted.
Independent measurement of the detector counting rate as a function of the background reveals a linear QE reduction with the external noise (dashed line).

 The dependencies presented in Fig. 2(a,b) were measured for the idler radiation wavelength equal to 780 nm. In our last experiment we measured the QE as a function of the wavelength; the result is shown in Fig. 2(c). Each point was measured at a different orientation of the crystal to provide the corresponding wavelength in the calibrated channel. For the Klyshko method, a set of interference filters was used to restrict the spectrum in the reference channel; for the difference-signal method, $D_s$ and $D_i$ were adjusted to
satisfy~(\ref{pm}). The results demonstrate a good agreement between both methods, as well as with
the data in the datasheet~\cite{datasheet}.
\begin{figure}
[h]\centerline{\includegraphics[width=6.5cm]{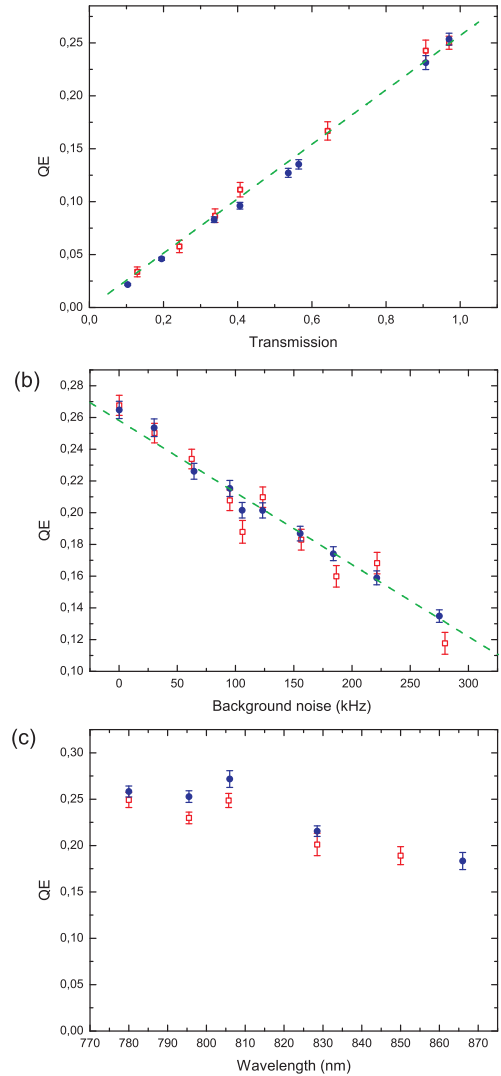}}
\caption{Experimental results. QE measured
by the difference-signal method (empty squares) and the coincidence-counting one (filled circles) versus the channel transmission
(a), the background noise (b), and the wavelength (c). Dashed line: theoretical dependence.}
\end{figure}

In conclusion, we have tested the new absolute calibration method based on the measurement of the difference-signal variance versus the traditional (Klyshko) one. The good agreement between the results verifies the validity of the new method. It also demonstrates the applicability of the new method to single-photon detectors, after taking into account the dead-time effect.

This work was supported in part by the Russian Foundation for Basic Research, grants \#\# 10-02-00202. T. Sh. Iskhakov acknowledges the support from Alexander von Humboldt Foundation. We are grateful to A. L. Chekhov for the help in calculations.

\end{document}